\documentclass[aps,bibnotes,twocolumn,showpacs, showkeys, superscriptaddress,preprintnumbers,amsmath,amssymb,lengthcheck,pra]{revtex4-1}
\usepackage{graphicx}% Include figure files
\usepackage{dcolumn}% Align table columns on decimal point
\usepackage{bm}% bold math
\usepackage{color}
\begin{document}

\title{Optical Kerr Spatio-Temporal Dark-Lump Dynamics of Hydrodynamic Origin}

%\author{Fabio Baronio} \email{fabio.baronio@unibs.it} 
%\affiliation{Dipartimento di Ingegneria dell'Informazione, Universit\`a di Brescia, \\
%Via Branze 38, 25123 Brescia, Italy}
%\author{Matteo Conforti}
%\affiliation{PhLAM/IRCICA UMR 8523/USR 3380,  CNRS-Universit\'e Lille 1,  F-59655 Villeneuve d'Ascq, France}
%\author{Shihua Chen}
%\affiliation{Department of Physics, Southeast University, Nanjing 211189,China}
%\author{Philippe Grelu}
%\affiliation{Laboratoire Interdisciplinaire Carnot de Bourgogne, UMR 6303 CNRS-Universit\'e de Bourgogne, BP 47870 Dijon Cedex 21078, France}
%\author{Stefan Wabnitz}
%\affiliation{Dipartimento di Ingegneria dell'Informazione, Universit\`a di Brescia, \\
%Via Branze 38, 25123 Brescia, Italy}

\author{Fabio Baronio} 
\email{fabio.baronio@unibs.it}
\author{Stefan Wabnitz} 
\affiliation{INO CNR and Dipartimento di Ingegneria dell'Informazione, Universit\`a di Brescia, Via Branze 38, 25123 Brescia, Italy}

\author{Yuji Kodama} 
\email{kodama@math.ohio-state.edu}
\affiliation{Department of Mathematics, Ohio State University, Columbus, OH 43210, USA}

\begin{abstract}
There is considerable fundamental and applicative interest in obtaining non-diffractive and non-dispersive spatio-temporal localized 
wave packets propagating in optical cubic nonlinear or Kerr media. 
Here, we analytically predict the existence of a novel family of
spatio-temporal dark lump solitary wave solutions of the (2+1)D nonlinear Schr\"odinger equation. 
Dark lumps represent multi-dimensional holes of light on a continuous wave background. We analytically derive the dark lumps from 
the hydrodynamic exact soliton solutions of the (2+1)D shallow water 
Kadomtsev-Petviashvili model,  inheriting their {complex interaction properties}. 
This finding opens a novel path for the excitation and control of optical multidimensional extreme wave phenomena of hydrodynamic
footprint.

%The hydrodynamic-like wave dynamics could be of great
%interest in a variety of complex optical systems.
%
%
%
\end{abstract}

%\ocis{190.0190, 190.3100, 190.4223, 190.5530}% REPLACE WITH CORRECT OCIS CODES FOR YOUR ARTICLE
                          %% NOTE: \ocis{} IS ALIASED TO \pacs{} BUT MUST
                          %% FORMAT THE TERMS CORRECTLY FOR EACH JOURNAL
													

\maketitle

%\section{Introduction}
\textit{Introduction.--}
The propagation of intense, ultra-short pulses of electromagnetic radiation in a nonlinear medium is a 
multi-dimensional phenomenon, leading to complex spatio-temporal behavior.
Pulse dynamics is influenced by the interplay of various physical
mechanisms: the most important among them being diffraction, material dispersion and 
nonlinear response \cite{boyd08}. 
Motivated by the strong applicative interest in the generation of high-intensity femtosecond pulses, 
a significant research activity on spatio-temporal light pulse propagation has been carried out over the
past decades. 
%
%The propagation of intense ultrashort light pulses has received a lot of attention 
%over the past decades. It is influenced by the interplay of variuos physical
%mechanisms, the most important being diffraction, material dispersion, and the 
%nonlinear Kerr effect \cite{boyd08}. 
%
Since the $1990$s, theoretical and experimental studies of the self-focusing behavior
of intense ultrashort pulses \cite{silber90,roth92a,ranka96,gaeta98,didda99,
wise99, suda00,eise01,berge01,baro02,baro04} have indicated that spatial and temporal degrees of freedom cannot be treated separately. 
When the three length scales naturally associated with diffraction, dispersion, and 
nonlinearity become comparable, the most intriguing consequence of space-time coupling 
is the possibility to form a non-diffractive and non-dispersive localized wave packet, 
namely, a spatiotemporal soliton or light bullet \cite{silber90}. 
A strict constraint for the formation of light bullets is that the nonlinear phase changes
counteract both the linear wave-front curvature and the dispersion-induced chirp, thus leading to 
space-time focusing \cite{wise99,eise01}. Vice versa, normal dispersion rules out the possibility to
generate bullet-type spatio-temporal localized wave packets. In this regime, qualitatively
different behaviors such as temporal splitting and spectral breaking have been observed
\cite{roth92a,ranka96,gaeta98,didda99}.
In the $2000$'s, theoretical and experimental studies have demonstrated that non-diffractive 
and non-dispersive localized wave packets also exist within the normal dispersion regime, in the form
of the so-called nonlinear X waves, or X-wave solitons \cite{conti03,ditra03,faccio06}.

Defeating the natural spatio-temporal spreading of wave packets is a challenging and 
universal task, appearing in any physical context that involves wave propagation phenomena. Ideal particle-like 
behavior of wave packets is demanded in a variety of applications, such as: microscopy, tomography, 
laser-induced particle acceleration, ultrasound medical diagnostics, Bose-Einstein condensation,
volume optical-data storage, optical interconnects, and those encompassing long-distance or 
high-resolution signal transmission.

In this Letter, we contribute to the field of non-diffractive and non-dispersive
spatio-temporal localized wave packets in cubic nonlinear (or Kerr) optical media,
by predicting the existence and the interactions of dark lump solitary wave solutions of the (2+1)D nonlinear 
Schr\"odinger equation (NLSE). 
The key point  of our approach consists in that we are able to derive the conditions for optical dark lump solitary 
waves existence, and analytically describe their shape and interactions, from the exact soliton 
solutions of the (2+1)D Kadomtsev-Petviashvili (KP) equation \cite{KP70}. 
In hydrodynamics, the KP equation describes weakly dispersive and small amplitude 
water wave propagation in a (2+1)D framework, 
in the so-called shallow water regime (see e.g. \cite{mil77,as81,kod10,kod11}). 
Our results recall and extend the connection between nonlinear wave propagation in optics and 
hydrodynamics, that was established in the $1990$'s to describe optical 
instabilities, dark stripe and vortex solitons in Kerr media \cite{turi88,law92,peli95,kivs96,malo98,kod95}.

Our treatment below goes as follows. We give first the essential
transformations that permit to construct dark solitary waves
of the optical (2+1)D NLSE, starting from exact multi-lump 
solutions of the KPI equation. 
We consider the propagation of dark-lump solitary
waves in the anomalous dispersion and self-defocusing regime. 
Then, we highlight {complex dark lumps' interactions} of the (2+1)D 
NLSE, that surprisingly mimic the behavior of multi-lump solutions of the KPI equation.
To conclude, we briefly discuss the conditions for the experimental observation of 
 dark lumps in nonlinear optics.

\textit{Optical NLSE solitary waves of hydrodynamic KP origin.--}
The dimensionless time-dependent paraxial wave equation in cubic Kerr media, 
in the presence of group-velocity dispersion, and limiting diffraction to one
dimension, reads as \cite{conti03}:
\begin{equation}\label{2DNLS}
iu_z+\frac{\alpha}{2}u_{tt}+\frac{\beta}{2}u_{yy}+\gamma |u|^2 u=0,
\end{equation}
where $u(t,y,z)$ represents the complex wave envelope; $t,y$
represent temporal and spatial transverse coordinates, respectively, and $z$ is the longitudinal propagation coordinate. 
Each subscripted variable in Eq. (\ref{2DNLS}) stands for
partial differentiation. $\alpha, \beta > 0, \gamma$ are real 
constants that represent the effect of dispersion, diffraction and Kerr nonlinearity, respectively.
%Here, we consider $\alpha < 0$, $\beta >0$ and $\gamma > 0$ 
%(i.e., normal dispersion, and focusing nonlinearity).
Of course, Eq. (\ref{2DNLS}) may also describe (2+1)D spatial
dynamics in cubic Kerr media, neglecting group-velocity
dispersion; in this case $t,y$
represent the spatial transverse coordinates, 
and $z$ the longitudinal propagation coordinate;
moreover $\alpha = \beta > 0$.

 %Eq. (\ref{2DNLS}) has been normalized in a way such that
%$\alpha < 0$, $\beta >0$ and $\gamma > 0$ (i.e., normal dispersion, and
%focusing nonlinearity).

Writing $u=\sqrt{\rho}\exp(i\theta)$, and substituting in Eq. (\ref{2DNLS}), we obtain for
the imaginary and real parts of the field the following system of equations for $(\rho,\theta)$,
\begin{align}
\rho_z+&\alpha\left(\rho\theta_t\right)_t+\beta\left(\rho\theta_y\right)_y  = 0, \nonumber\\[0.5ex]
\theta_z-&\gamma\rho+\frac{\alpha}{2}\left(\theta_t^2+\frac{1}{4\rho^2}\rho_t^2-\frac{1}{2\rho}\rho_{tt}\right) + \nonumber \\[0.5ex]
&~~+\frac{\beta}{2}\left(\theta_y^2+\frac{1}{4\rho^2}\rho_y^2-\frac{1}{2\rho}\rho_{yy}\right)=0.
\label{2DNLSb}
\end{align}
Let us consider now small corrections to the stationary continuous wave (CW) background solutions of Eqs. (\ref{2DNLSb}), and set 
\begin{equation}
\rho=\rho_0+\eta, \qquad \theta=\gamma\rho_0 z+\phi,
\end{equation}
 with constant $\rho_0$. 
 With a small positive parameter $0<\epsilon \ll 1$,
we assume the following scaling 
$\eta\,\sim\,\phi_z\,\sim\,\phi_t\,\sim\,\mathcal{O}(\epsilon)$, $\partial_t\,\sim\,\partial_z\,\sim\,\mathcal{O}(\epsilon^{1/2})$, $\partial_y\,\sim\,\mathcal{O}(\epsilon)$. Then we obtain from Eqs. (\ref{2DNLSb}) 
\begin{align}
\eta_z+\rho_0(\alpha\phi_{tt}+\beta\phi_{yy})+\alpha(\eta\phi_t)_t&=\mathcal{O}(\epsilon^{7/2}), \nonumber\\
\phi_z-\gamma\,\eta+\frac{\alpha}{2}\left(\phi_t^2-\frac{1}{2\rho_0}\eta_{tt}\right) &=\mathcal{O}({\epsilon^3}).\label{2DNLSd}
\end{align}
Introducing the coordinates $\tau=t-c_0 z,$ $\upsilon =y,$ $\varsigma=z$ ($c_0=\sqrt{-\gamma\alpha\rho_0}$),
and noting that $\partial_\varsigma\sim\mathcal{O}(\epsilon^{3/2})$, from Eqs. (\ref{2DNLSd})
we have
\begin{align}
-c_0\eta_{\tau}+\eta_{\varsigma}+\rho_0\alpha\phi_{\tau\tau}+\rho_0\beta\phi_{\upsilon\upsilon}+\alpha(\eta\phi_\tau)_\tau&=\mathcal{O}(\epsilon^{7/2}) \nonumber\\[0.5ex]
-c_0\phi_{\tau}+\phi_{\varsigma}-\gamma\eta+\frac{\alpha}{2}\left(\phi_\tau^2-\frac{1}{2\rho_0}\eta_{\tau\tau}\right) &=\mathcal{O}({\epsilon^3}). \label{2DNLSf}
\end{align}
From the second of Eqs. (\ref{2DNLSf}),  we obtain $\eta=-\frac{c_0}{\gamma} \phi_{\tau} +$ \textit{(higher order terms)};
iterating to find the higher order terms, we obtain
\begin{align}\label{etaphi}
\eta=\frac{1}{\gamma}\left(-c_0\phi_{\tau}+\phi_\varsigma+\frac{\alpha}{2}\phi_\tau^2-\frac{\alpha^2}{4c_0}\phi_{\tau\tau\tau}\right)+\mathcal{O}(\epsilon^3).
\end{align}
By inserting (\ref{etaphi}) in the first of Eqs. (\ref{2DNLSf}),  we have
\begin{align}\label{PKP}
\phi_{\tau\varsigma}+\frac{3\alpha}{4}(\phi_{\tau}^2)_\tau-\frac{\alpha^2}{8c_0}\phi_{\tau\tau\tau\tau}+\frac{c_0\beta}{2\alpha}\phi_{\upsilon\upsilon}=\mathcal{O}(\epsilon^3).
\end{align}
Eq. (\ref{PKP}) is known as the potential KP equation \cite{kod10}.
In fact, from Eq. (\ref{PKP}) we obtain the evolution equation for $\eta$, namely, we have
the KP equation at the leading order,
\begin{equation}\label{KP}
\left(-\eta_\varsigma+\frac{3\alpha\gamma}{2c_0}\eta\eta_\tau+\frac{\alpha^2}{8c_0}\eta_{\tau\tau\tau}\right)_\tau-\frac{c_0\beta}{2\alpha}\eta_{\upsilon\upsilon}=0.
\end{equation}
Notice that, in the case $\alpha > 0$, $\beta>0, \gamma<0$, we have the KP I case,
and when $\alpha <0$, $\gamma>0, \beta>0$, the KP II case. 

Therefore, we underline that the optical NLSE solution $u(t,y,z)$ of hydrodynamic
KP solution origin [$\eta(\tau,\upsilon,\varsigma), \phi(\tau,\upsilon,\varsigma)$]
with $\tau=t-c_0z, \upsilon=y$ and $\varsigma=z$
can be written as:
\begin{align}\label{NLSEKP}
u(t,y,z)=\sqrt{\rho_0+\eta(\tau,\upsilon,\varsigma)} \ \ e^{i(\gamma\rho_0z +\phi(\tau,\upsilon,\varsigma))}.
%u(t,y,z)=\sqrt{\rho_0+\eta(t=\tau,y=\upsilon,z=\varsigma)} \ \ e^{i(\gamma \rho_0 z +\phi(t=\tau,y=\upsilon,z=\varsigma))}.
\end{align}
%with $\tau=t-c_0 z,$ $\upsilon =y,$ $\varsigma=z$.
%
%with $\rho_0$ intensity level background, $\theta_0=\gamma \rho_0 z$, 
%$\tau=t-c_0 z,$ $\upsilon =y,$ $\varsigma=z$.
%
%
%Writing $u=\sqrt{\rho}\exp(i\theta)$, we have
%\begin{align*}
%u_z&=\left(\frac{1}{2\sqrt{\rho}}\rho_z+i\sqrt{\rho}\,\theta_z\right)e^{i\theta}\\
%u_{tt}&=\left(\frac{1}{2\sqrt{\rho}}\rho_{tt}-\frac{1}{4\rho^{3/2}}\rho_t^2-\sqrt{\rho}\theta_t^2
%+i\left(\frac{1}{\sqrt{\rho}}\rho_t\theta_t+\sqrt{\rho}\theta_{tt}\right)\right)
%\end{align*}
%%
%
%
%
%
%

In the following, we focus our attention on the anomalous dispersion and self-defocusing regime
($\alpha > 0$, $\beta>0, \gamma<0$), which leads to the KPI case.
The normal dispersion and self-focusing regime ($\alpha <0$, $\gamma>0, \beta>0$),
which leads to the KPII case, will be analyzed in a future work. 
Without loss of generality, we set the following constraints to the coefficients 
of Eq. (\ref{2DNLS}), $\alpha=4 \sqrt{2}, \beta= 6\sqrt{2}, \gamma=-2\sqrt{2}$;
moreover, we fix $\rho_0=1$.
%, so that $c_0=4$. 
Note that, with the previous relations among its coefficients, the Eq. (\ref{KP}) reduces to the 
standard KPI form \cite{spKP}.

{\textit{Single NLSE dark lump solution of KPI origin.--}}
At first, we proceed to verify numerically the existence of (2+1)D NLSE dark-lump solitary wave,
which is predicted by the KPI through Eq.(\ref{NLSEKP}) (see e.g. \cite{mana77,as81}
for the lump solutions of KPI).
In our numerics, the input dark solitary wave envelope at $z=0$ is given by the expression
$u(t,y,0) = \sqrt{1+\eta(\tau,\upsilon,0)} \exp{\left[i \phi(\tau,\upsilon,0)\right]}$ with
$\tau=t$ and $\upsilon=y$,
where $\eta$ is a bright lump solution of the KPI equation (\ref{KP}), and $\phi_\tau=-(\gamma/c_0) \, \eta$.

When considering the small amplitude regime ($\epsilon \ll 1$), a form of KP lump-soliton solution of Eq. (\ref{KP}) can be expressed as 
$\eta (\tau, \upsilon, \varsigma) = -4 [ \epsilon^ {-1}-(\tau-3 \epsilon \varsigma)^2+\epsilon \upsilon^2   ] /  [ \epsilon^ {-1}+(\tau-3 \epsilon \varsigma)^2+\epsilon\upsilon^2   ]^2$.
The parameter $\epsilon$ rules the amplitude/width and velocity properties of the KP lump soliton.
The lump peak amplitude in the $(\varsigma,\upsilon)$ plane is $-4 \epsilon$; the velocity in the  $\tau$-direction
 is $3\epsilon$. Moreover,  $\phi(\tau, \upsilon, \varsigma)= 2 \sqrt{2} \epsilon (\tau-3 \epsilon \varsigma)/ [1+ \epsilon (\tau-3 \epsilon \varsigma)+\epsilon^2 \upsilon^2].$

\begin{figure}[h!]
\includegraphics[width=7cm]{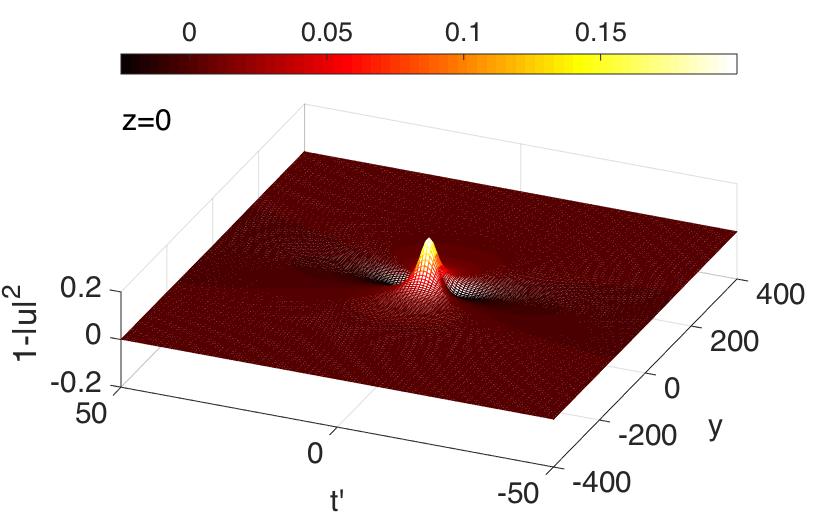}
\includegraphics[width=7cm]{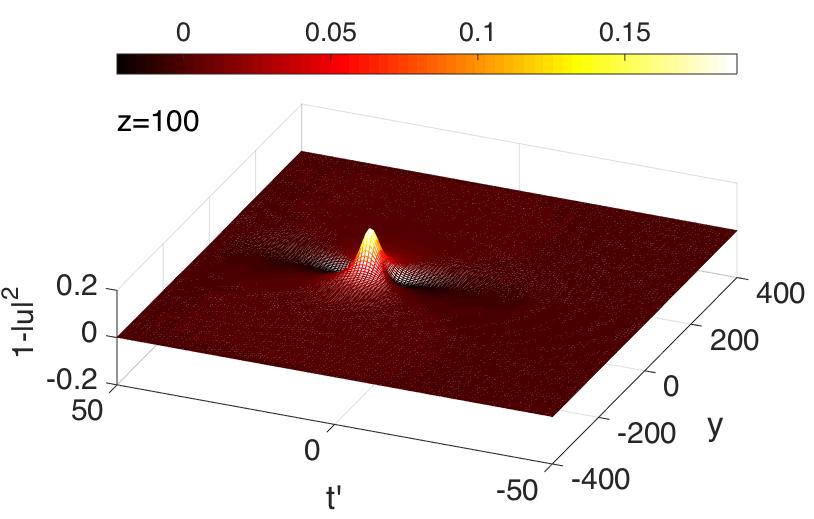}
\caption{\label{f1} Numerical spatio-temporal dark-lump NLSE envelope intensity distribution
$1-|u|^2$, shown in the $y$-$t'$ plane with $t'=t-c_0z$, at $z=0$, and $z=100$. Here, $\epsilon=0.05$. 
}
\end{figure}

Figure \ref{f1} shows the numerical spatio-temporal envelope intensity profile
$1-|u|^2$ of  a NLSE dark lump solitary wave in the $y$-$t'$ plane ($t'=t-c_0 z$), 
 at the input $z=0$ and after the propagation distance $z=100$, for $\epsilon=0.05$. 
In the numerics, the initial dark NLSE profile, of KPI lump, propagates stably in the $z$-direction,
with virtually negligible emission of dispersive waves,
 with the predicted velocity $c_0+3\epsilon$,
and intensity dip of $4 \epsilon$. 
Thus, the predicted theoretical dark lump solitary waves of Eq. (\ref{NLSEKP}) 
are well confirmed by numerical simulations. 

%\onecolumngrid

%

%\twocolumngrid
%
%\begin{figure}
%\includegraphics[width=7cm]{err.eps}
%\includegraphics[width=8cm]{f1.eps}
%\caption{\label{f2} Error function $e$ [\%] versus $\epsilon$ values. }
%\end{figure}
%

Remarkably, our numerical studies have shown that the previously described NLSE--KP mapping works well also for 
values of $\epsilon$ which lead to strong perturbations of the stationary CW backgrounds (this will be reported elsewhere).

\textit{Elastic interaction of two single NLSE dark lumps.--}
Next, we consider the interaction of (2+1)D NLSE optical dark lumps 
based on the multi-lump solutions of the KPI equation (see e.g. \cite{mana77,as81,grim04}).
%
%In our numerics, the input dark solitary wave envelope Eq. (\ref{NLSEKP}) is given by the expression
%$u(t,y,z=0) = \sqrt{1+\eta(\tau=t,\upsilon=y,\varsigma=0)} \exp{\left[i \phi(\tau=t,\upsilon=y,\varsigma=0)\right]}$, 
%where $\eta$ is the sum of N peaked lump-solitons of Eq. (\ref{KP}).

A formula for the exact multi-lump solution is available (see e.g. \cite{mana77,as81,grim04}).
However, when the lumps are well-separated, a simple sum of those single lump solutions
give a good approximation of the exact $N$ lump solution, that is,
$\eta (\tau, \upsilon, \varsigma)\approx \sum\limits_{i=1}^N -4 [ \epsilon_i^ {-1}-(\tau_i-3 \epsilon_i \varsigma)^2+\epsilon_i \upsilon_i^2   ] /  [ \epsilon_i^ {-1}+(\tau_i-3 \epsilon_i \varsigma)^2+\epsilon_i\upsilon_i^2   ]^2$, where $\epsilon_i$ rules the amplitude/width and velocity properties of the $i$-lump soliton,  $\tau_i=\tau-\tau_{0i}, \upsilon_i=\upsilon-\upsilon_{0i}$
define the $i$-lump's location.

\begin{figure}[b]
\includegraphics[width=6.5cm]{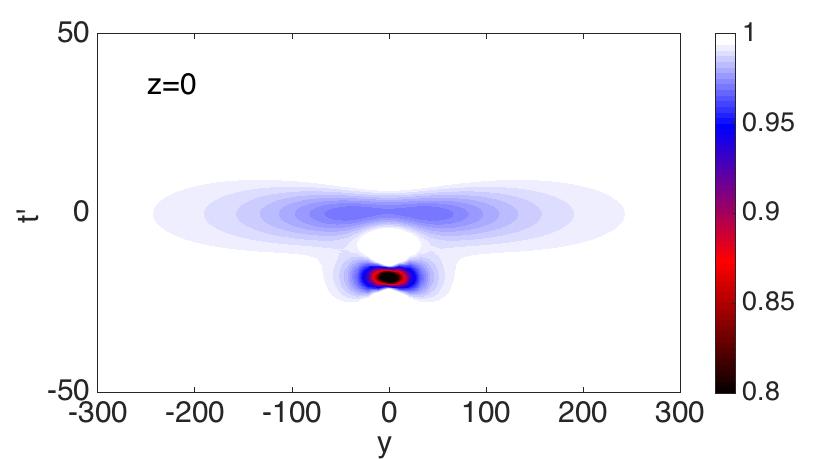}
\includegraphics[width=6.5cm]{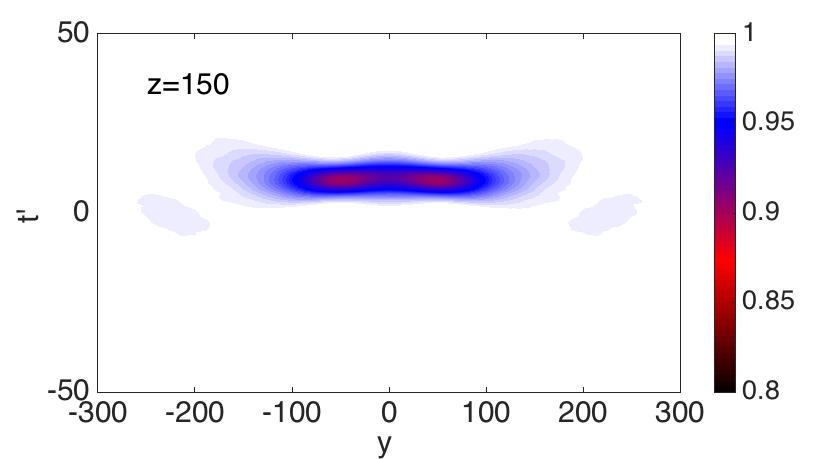}
\includegraphics[width=6.5cm]{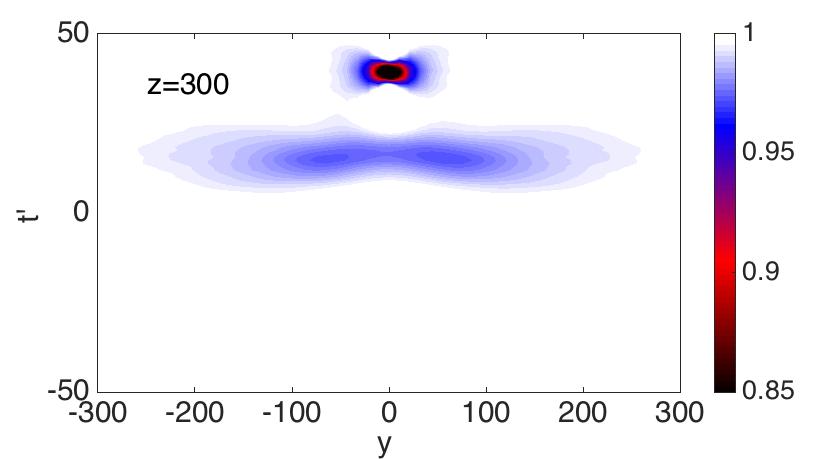}
\caption{\label{felastic} Numerical spatio-temporal NLSE envelope intensity distribution
$|u|^2$, in the $y$-$t'$ plane, showing the interaction of two dark-lumps ($N=2$), at the input $z=0$, 
at $z=150$, and $z=300$. 
Here, $\epsilon_1=0.09, \tau_{01}=-20, \upsilon_{01}=0$ , $\epsilon_2=0.01, \tau_{02}=0, \upsilon_{02}=0$.
}
\end{figure}

Figure \ref{felastic} shows the initial spatio-temporal envelope intensity profile
$|u|^2$, of  NLSE dark lumps for $N=2$ in the $y$-$t'$ plane, 
along with the numerically computed profiles after propagation distances $z=150$, 
and $z=300$.
Here, $\epsilon_1=0.09, \tau_{01}=-20, \upsilon_{01}=0$, $\epsilon_2=0.01, \tau_{02}=0, \upsilon_{02}=0$.
One can see from Fig. \ref{felastic} that two dark-lumps with different amplitudes
shows an \emph{elastic} interaction:
the tall lump approaches the small one along the $t'$-axis, then they interact
and generate a wave form with two separate peaks in the $y$-direction.
After the interaction, the tall soliton is in front of the small one and those
lumps keep their profiles.

%\onecolumngrid

%\twocolumngrid

%
%
%

\textit{Abnormal scattering of NLSE dark lump solution.--}
At last, we remark that the KPI equation admits another type of lump solutions which have several peaks
with the same amplitude in the asymptotic stages $|z|\gg 0$ (see e.g. \cite{john78,ablo00}).
Following \cite{ablo00}, we call such lump solution \emph{multi-pole lump}.
Here we demonstrate that $(2+1)$D NLSE can also support such lump solution.
We consider multi-pole lump solution with two peaks, which is expressed as \cite{ablo00}: $\eta (\tau, \upsilon, \varsigma) = -2 \partial_\tau ^2 {\rm log} F$,
where $F=|f_1^2+i f_2 +f_1/ \epsilon+1 /2 \epsilon^ {2}|^2+ |f_1+1/ \epsilon|^2 / 2\epsilon^2+1/4\epsilon^ {4}$,
and $f_1=\tau_1+2i \epsilon \upsilon_1-12 \epsilon^2 \varsigma_1 +\delta_1$, $f_2=-2 \upsilon-24i \epsilon \varsigma+\delta_2$.
$\tau_1=\tau-\tau_{0}, \upsilon_1=\upsilon-\upsilon_{0},   \varsigma_1=\varsigma-\varsigma_0$
define the dislocation; $\delta_1, \delta_2$ are arbitrary complex parameters.

\begin{figure}
\includegraphics[width=5.5cm]{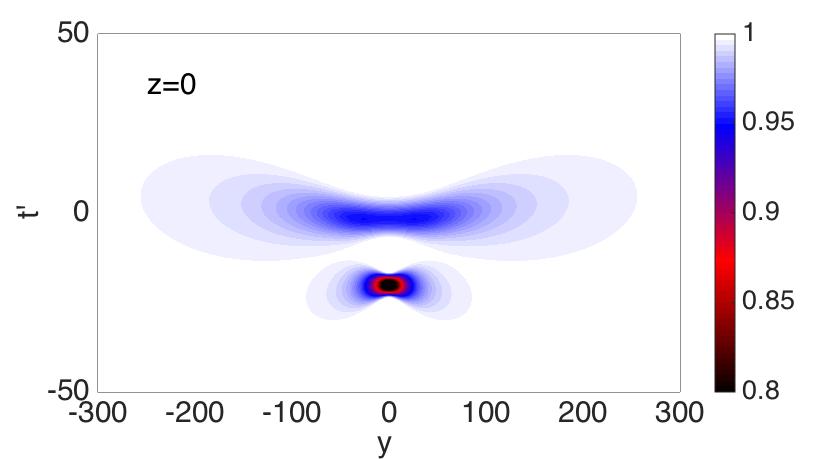}
\includegraphics[width=5.5cm]{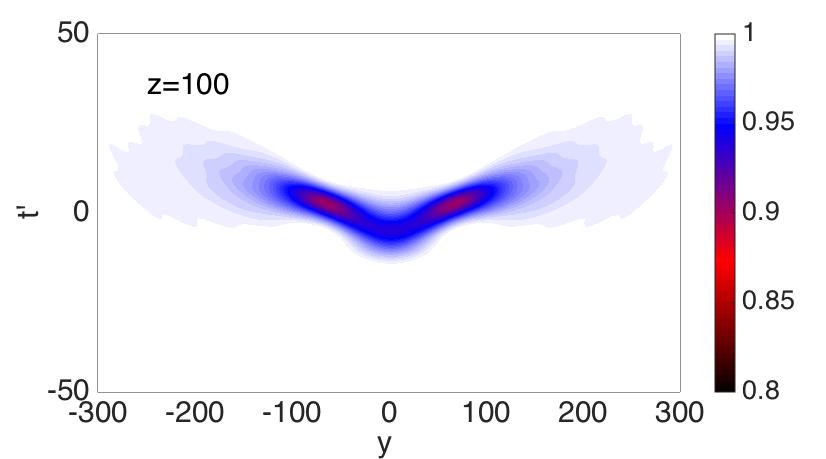}
\includegraphics[width=5.5cm]{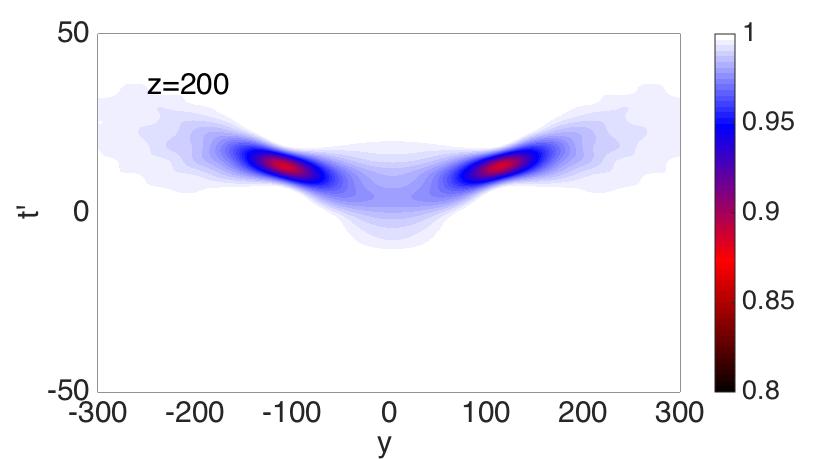}
\includegraphics[width=5.5cm]{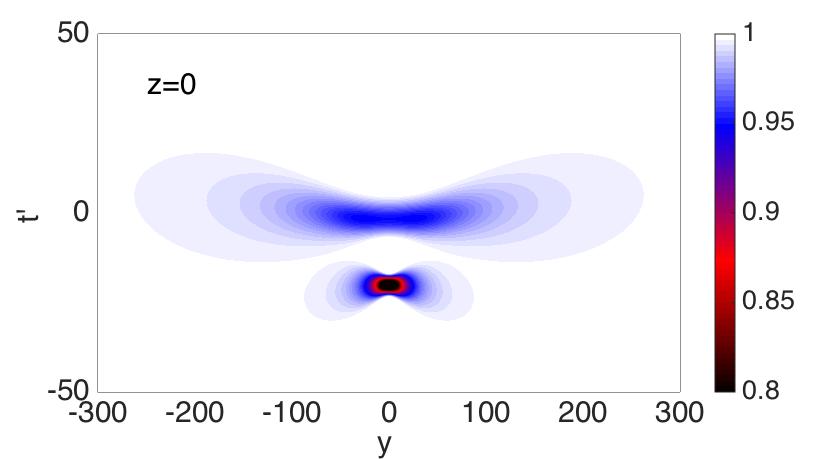}
\includegraphics[width=5.5cm]{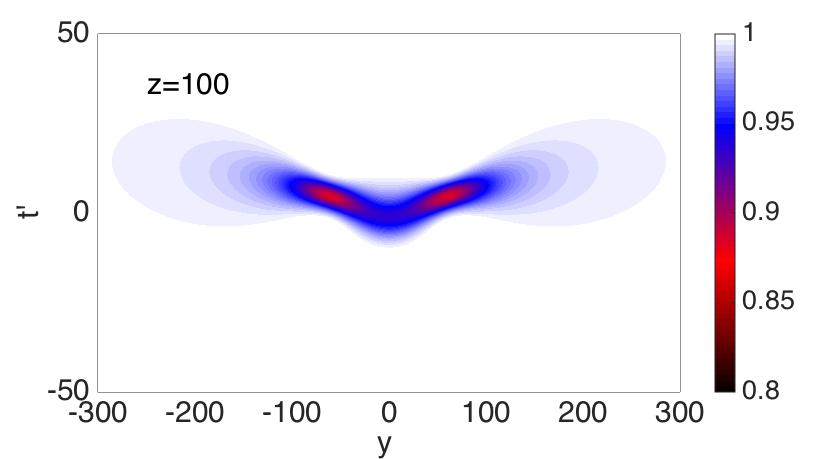}
\includegraphics[width=5.5cm]{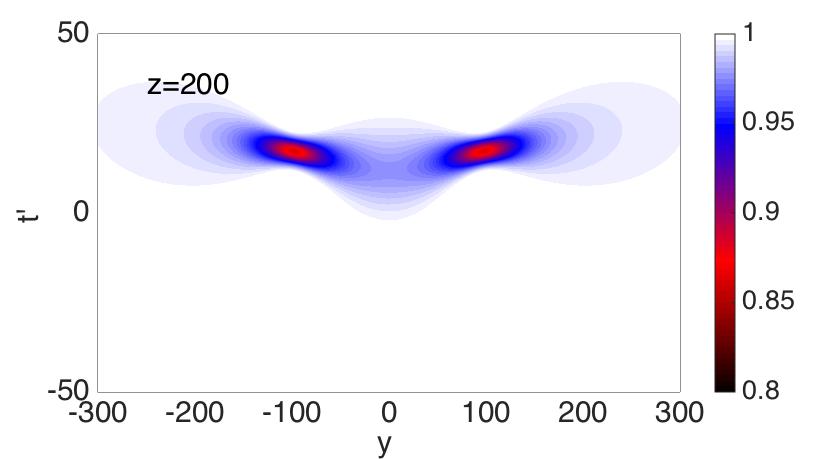}
\caption{\label{f2} Spatio-temporal NLSE envelope intensity distribution
$|u|^2$, in the $y$-$t'$ plane, showing anomalous scattering of dark waves, at $z=0$, 
at $z=100$ and $z=200$. 
Top, numerical simulations; bottom theoretical prediction Eq. (\ref{NLSEKP})
with KPI multi-pole lump solution.
Here, $\epsilon=0.1$, $\tau_0=0, \upsilon_0=0, \varsigma_0=-50, \delta_1=0, \delta_2=0$ 
}
\end{figure}

Figure \ref{f2} (top) shows the initial spatio-temporal envelope intensity profile
$|u|^2$ of a two peaked NLSE dark lump in the $y-t'$ plane, 
along with the numerically computed profiles after propagation distances $z=100$, 
and $z=200$, for $\epsilon=0.1$ ($\tau_0=0, \upsilon_0=0, \varsigma_0=-50, \delta_1=0, \delta_2=0$). 
In particular, Fig. \ref{f2} depicts the scattering interaction of the two-peaked waves:
two dark lumps approach each other along the $t'$-axis, interact, and 
recede along the $y$-axis.
These solutions exhibit anomalous (nonzero deflection angles) scattering due to multi-pole 
structure in the wave function of the inverse scattering problem.
We remark that the numerical result of NLSE dynamics 
is in an excellent agreement with analytical dark solitary solution Eq. (\ref{NLSEKP})
with KPI multi-pole lump solution, 
as seen in Fig. \ref{f2} (bottom).  

%Remarkably, the observed dark solitary NLSE interaction fully inherits the 
%hydrodynamic KP bright soliton interaction behavior \cite{ablo00}. 

\onecolumngrid

%\twocolumngrid

\medskip

\vskip 1cm

\twocolumngrid

\textit{Experiments in Optics.--}
Let us briefly discuss a possible experimental setting in nonlinear optics
for the observation of cubic spatio-temporal solitary wave dynamics of hydrodynamic origin. 
As to (2+1)D spatio-temporal dynamics,  one may consider optical propagation in a planar glass 
waveguide (e.g., see the experimental set-up   of Ref. \cite{eise01}), or  a quadratic lithium
niobate planar waveguides, in the regime of high  phase-mismatch, which mimics an effective 
Kerr nonlinear regime (e.g., see the experimental set-up of Ref. \cite{baro02}). 
As to (2+1)D spatial dynamics, one may consider a CW Ti:sapphire laser and a nonlinear medium 
composed of atomic-rubidium vapor (e.g., see the experimental set-up   of Ref. \cite{kivs96}),
or a bulk quadratic lithium
niobate crystal, in the regime of high  phase-mismatch (e.g., see the experimental set-up of Ref. \cite{baro04, krupa15}).

The excitation of spatio-temporal dark lump solitary waves from non-ideal input conditions is a relevant problem for the experiments, 
and it will be the subject of further investigations. 

As a final remark, note that the well known modulation instability of plane waves 
\cite{yuen,conti03}, or conical emission, in general may emerge in the (2+1)D NLSE 
scenario. In the case we have considered, that is anomalous dispersion and self-defocusing regime
($\alpha > 0$, $\beta>0, \gamma<0$, the NLSE- KPI correspondence), MI
is absent. On the other hand, when considering the normal dispersion and self-focusing regime 
($\alpha <0$, $\gamma>0, \beta>0$, the NLSE - KPII correspondence) 
MI plays a crucial competing role.
In fact, the modulation instability of the CW background may compete and ultimately spoil, 
for sufficiently long propagation distances, the propagation and
interaction of dark solitary waves in (2+1)D NLSE propagation.

\textit{Conclusions.--}
We have analytically predicted a new class of dark solitary wave solutions 
that describe non-diffractive and non-dispersive spatio-temporal localized 
wave packets propagating in optical Kerr media.
We numerically confirmed the existence, stability, and peculiar
 elastic and anomalous scattering interactions of dark-lump solitary waves of the (2+1)D NLSE. 
The key novel property of these solutions is that their existence 
and interactions are inherited from the hydrodynamic soliton 
solutions of the well known KP equation.
Our findings open a new avenue for research in spatio-temporal extreme nonlinear optics. 
Given that deterministic rogue and shock wave solutions, so far, have been essentially restricted to 
(1+1)D models \cite{kib10,bar12,ono13,bar14,rand14,conf14,kibl15}, 
multidimensional spatio-temporal nonlinear waves would lead to a 
substantial qualitative enrichment of the landscape of extreme wave phenomena.

%\section{Conclusions}
%In our opinion, all the shallow water 
%KP interactions \cite{kod03,kod10} would be reproduced in the NLSE optical domain
%(a complete treatment of NLSE higher order solitary solution of KP origin will be reported
%elsewhere).   
%These results provide a clear evidence that the mapping Eq. (\ref{NLSEKP}) permits to transfer 
%theoretical and experimental phenomenologies of the hydrodynamic KP dynamics (i.e., soliton interactions, 
%rogue waves, tsunamis, etc.) into the realm of multidimensional nonlinear optics. 

The present research was supported by the Italian
Ministry of University and Research (MIUR, Project No. 2012BFNWZ2).
The research of YK is partially supported by NSF grant, DMS-1410267.
%\clearpage

\end{document}